\documentclass{emulateapj}
\usepackage{apjfonts}
\usepackage{amsmath}
\usepackage{graphicx}
\usepackage{dcolumn}

\newcounter{ichi}
\newcounter{ni}
\newcounter{san}
\newcounter{yon}
\makeatletter
\newsavebox{\@parc@ption}
\def\parcaption#1{%
\sbox{\@parc@ption}{\shortstack[l]{#1}}%
>\setbox\@tempboxa\hbox{\csname fnum@\@captype\endcsname}%
\@tempdima\columnwidth \advance\@tempdima-\wd\@tempboxa
\@tempdimb\@tempdima 
\ifdim\wd\@parc@ption>\@tempdimb \@tempdima\@tempdimb
\else\@tempdima\wd\@parc@ption\fi
\sbox{\@tempboxa}{\parbox[t]{\@tempdima}{#1}}%
\caption{\usebox{\@tempboxa}}}
\makeatother

\slugcomment{}

\shorttitle{
TeV Synchrotron Pair Echoes and Haloes from Ultra-High-Energy Cosmic-Ray Accelerators
}
\shortauthors{Kohta Murase}

\begin{document}

\title{High-Energy Emission Induced by Ultra-High-Energy Photons as a Probe of Ultra-High-Energy Cosmic-Ray Accelerators Embedded in the Cosmic Web}

\author{Kohta Murase\altaffilmark{1}
}

\altaffiltext{1}{Department of Physics, Center for Cosmology and AstroParticle Physics, The Ohio State University, Columbus, OH 43210, USA}

\begin{abstract} 
The photomeson production in ultra-high-energy cosmic-ray (UHECR) accelerators such as $\gamma$-ray bursts and active galaxies may lead to ultra-high-energy (UHE) $\gamma$-ray emission.  We show that generation of UHE pairs in magnetized structured regions embedding the sources is inevitable, and accompanied $\gtrsim 0.1$~TeV synchrotron emission provides an important probe of UHECR acceleration. It would especially be relevant for powerful transient sources, and synchrotron pair echoes may be detected by future CTA via coordinated search for transients of duration $\sim 0.1-1$~yr for the structured regions with $\sim$~Mpc.  Detections will be useful for knowing structured extragalactic magnetic fields as well as properties of the sources.
\end{abstract}


\section{Introduction}
The origin of ultra-high-energy cosmic rays (UHECRs) has been a big mystery in astroparticle physics~\citep[see reviews, e.g.,][]{BS00,Der07,KO11}.  Despite recent observational progress by HiRes, the Pierre Auger Observatory (PAO) and the Telescope Array, interpretations of crucial information, anisotropy~\cite{PAO07,PAO10a,HiR10a,PAO11} and composition~\cite{PAO10b,HiR10b}, have not been settled and different arguments have been suggested~\citep[e.g.,][]{Der07,Gor+08,LW09,MT09,Zaw+09,BS11}.  
To identify sources or accelerators and investigate their physical mechanisms, other messengers, high-energy $\gamma$ rays and neutrinos, should be important.  Now, a km$^3$ neutrino detector, IceCube, was completed, and \textit{Fermi} and ground-based Cherenkov $\gamma$-ray detectors are operating.  
At TeV energies, the Cherenkov Telescope Array (CTA) with much better sensitivities~\cite{CTA10}, is being planned.  

Various candidates of the sources, including active galactic nuclei (AGN), $\gamma$-ray bursts (GRBs) and newly born magnetars, have been suggested~\cite[see reviews, e.g.,][and references therein]{Der07,KO11}.  Theoretically, $E \gtrsim {10}^{20}$~eV cosmic-ray accelerators must be powerful enough, and the magnetic luminosity of those relativistic outflow sources may satisfy $L_B \gtrsim {10}^{47.2}~{\rm erg}~{\rm s}^{-1}~ {(E_{20}/Z)}^2 \Gamma_1^2$~\cite{Bla00,FG09,LW09}, where $E={10}^{20}~{\rm eV}~E_{20}$ and $\Gamma = 10 \Gamma_1$ is the outflow Lorentz factor.  
This is even the case if UHECRs mainly consist of protons, which tempts us to consider transient sources like GRBs~\cite{LW09}.  In the AGN case, flaring activities have the advantage in explaining that the power of correlating AGN seems insufficient to produce UHECRs~\cite{Der+09,LW09,Zaw+09}.  

For charged cosmic rays, extragalactic magnetic fields (EGMFs) in structured regions embedding the sources and the Galactic magnetic field play crucial roles in causing deflections and significant time delays~\citep[e.g.,][]{Tak+06,KL08,MT09,KO11}.  On the other hand, photons and neutrinos, which are generated in the sources, can be more beamed and coincident with their activities.  For transients, due to the difficulty in revealing the sources with UHECRs, it is more favorable to detect photons and neutrinos.   

We study a characteristic signature of $E_p \gtrsim {10}^{20}$~eV cosmic-ray accelerators.  In powerful UHECR sources, one may expect ultra-high-energy (UHE) $\gamma$ rays as well as neutrinos produced via the $p \gamma$ reaction, which induce intergalactic cascades.  We show that a significant energy fraction should appear as synchrotron emission in magnetized structured regions, and this synchrotron pair echo is a crucial probe especially for powerful transients. 

\section{Production and fate of UHE photons}
If cosmic rays are accelerated up to ultra-high energies in GRBs and AGN, the $p \gamma$ reaction between protons and photons \textit{inside} the source should lead to hadronic $\gamma$ rays and neutrinos~\citep[e.g.,][]{WB97,RM98,AD01,AD03,Mur07,Mur09}.  Their spectra are calculated given proton and target photon spectra, and we use a (broken) power-law photon spectrum which is expected in the electron synchrotron emission mechanism: $dn/d\varepsilon \propto \varepsilon^{-\alpha}$.  Here $\varepsilon$ is the target photon energy in the comoving frame of the outflow with $\Gamma$ (while $\varepsilon_{\rm ob} \approx \Gamma \varepsilon$ is the energy in the observer frame). 
For GRBs and AGN, $\alpha \sim 1-1.5$ for $\varepsilon < \varepsilon^b$ and $\alpha \sim 2$ for $\varepsilon > \varepsilon^b$ are observed as typical values, where $\varepsilon^b$ is the break energy.  Then, using the rectangular approximation, the effective optical depth for the $p \gamma$ reaction is estimated to be~\citep[e.g.,][]{WB97,RM98,Mur+08}
\begin{equation}
f_{p \gamma} \approx \frac{t_{\rm dyn}}{t_{p \gamma}} \simeq \frac{2 \kappa_\Delta \sigma_\Delta}{1+\alpha} \frac{\Delta \bar{\varepsilon}_{\Delta}}{\bar{\varepsilon}_{\Delta}} 
\frac{L_{\gamma}^b}{4 \pi r \Gamma^2 c \varepsilon_{\rm ob}^b} {\left(\frac{E_p}{E_p^b} \right)}^{\alpha-1}, 
\end{equation}
where $t_{\rm dyn} \approx r/\Gamma c$ is the dynamical time, $t_{p \gamma}$ is the $p \gamma$ cooling time, $\sigma_\Delta \sim 5 \times {10}^{-28}~{\rm cm}^2$, $\kappa_\Delta \sim 0.2$, $\bar{\varepsilon}_{\Delta} \sim 0.3$~GeV, $\Delta \bar{\varepsilon}_{\Delta} \sim 0.2$~GeV, $L_\gamma^b$ is the luminosity at $\varepsilon_{\rm ob}^b$, $r$ is the emission radius, and $E_p^b \approx 0.5 \Gamma^2 m_p c^2 \bar{\varepsilon}_\Delta/\varepsilon_{\rm ob}^b$. 
For AGN with $L_\gamma^b={10}^{45.5}~{\rm erg/s}~$, $\varepsilon_{\rm ob}^b =10$~eV, $\Gamma=10$ and $r={10}^{17.5}$~cm, one has $f_{p \gamma} \simeq 0.88 \times {10}^{-3} {(E_p/1.6 \times {10}^{18}~{\rm eV})}^{0.5}$, while $f_{p \gamma} \simeq 0.022$ for typical high-luminosity (HL) GRBs with $L_\gamma^b = {10}^{51.5}~{\rm erg/s}$, $\varepsilon_{\rm ob}^b = 500$~keV, $\Gamma={10}^{2.5}$ and $r={10}^{14.5}$~cm though the multi-pion production enhances $f_{p \gamma}$ by a factor of $\sim 3$~\cite{Mur07,Mur08}. 
Hadronic $\gamma$ rays are generated via $\pi^0 \rightarrow 2 \gamma$, and their generation spectrum is roughly expressed as $E^2 \phi^{\rm gen} \propto f_{p \gamma} E_p^{2-p} \propto E^{1+\alpha-p}$ thanks to $\pi^0$'s short lifetime ($\tau_{\pi^0} =8.4 \times {10}^{-17}$~s).   
Then, cascades in the source happen at energies where $\gamma \gamma$ optical depth $\tau_{\gamma \gamma}$ is large.  On the other hand, UHE photons may also escape from synchrotron sources (e.g., GRBs and high-peaked BL Lacs) due to synchrotron self-absorption suppression below $\varepsilon_{\rm ob}^{\rm sa}$ in the target photon spectrum~\cite{Raz+04,LW07,Mur09}~\footnote{The emission zone of UHE photons cannot be too strongly magnetized to avoid 1$\gamma$ pair creation, i.e., $(E_\gamma/2\Gamma m_e c^2) (B_{\perp}^{\prime}/B_Q) \lesssim 1$ ($B^{\prime} \lesssim 2000~{\rm G}~\Gamma_{2.5} E_{\gamma,19}^{-1}$), implying sufficient large emission radii.  For GRBs, it can be realized in afterglow, and prompt emission when the target photon field exists above the acceleration region of UHE protons (that is plausible).}.  The escape fraction $f_{\rm esc}$ is estimated by $e^{-\tau_{\gamma \gamma}}$ for the instantaneous emission from a thin shell or $(1- e^{-\tau_{\gamma \gamma}})/\tau_{\gamma \gamma}$ in the emitting slab.  The escape can be easier when more detailed effects are included~\cite{Gra+08}.  

We hereafter consider such UHE photon sources.  The typical energy of $\gamma$ rays produced by the $p \gamma$ reaction is $E_\gamma \approx {10}^{19}~{\rm eV}~E_{p,20}$, so they can provide evidence of UHECR acceleration~\cite{Mur09}.  They are cascaded (or attenuated) in intergalactic space due to interactions with the cosmic infrared, microwave, and radio background (CIB/CMB/CRB).  Their interaction length for $\gamma \gamma$ pair creation is $\lambda_{\gamma \gamma} \sim 2.1~{\rm Mpc}~E_{\gamma,19} (10/\ln (4200 E_{\gamma,19}))$ (see Figure~1), and UHE pairs with $\gamma_e \sim 2 \times {10}^{13} E_{\gamma,19}$ are generated.

\begin{figure}[tb]
\includegraphics[width=\linewidth]{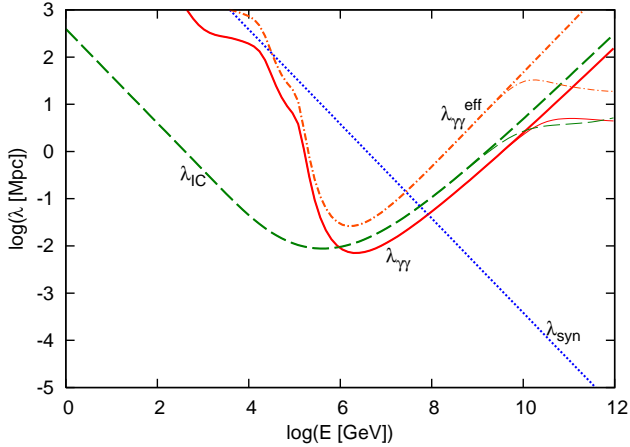}
\caption{
\small{The interaction length (solid curves) and effective loss length (dotted-dashed curves) of high-energy photons for the pair creation process, the energy loss length of electron-positron pairs for the inverse-Compton process (dashed curves), and the synchrotron cooling length for $B_{\rm EG}={10}^{1.5}$~nG (dotted curve).  Thick/thin curves represent cases without/with the CRB. 
}}
\end{figure} 

Our Universe has the large scale structure and this inhomogeneity is crucial for propagation and resulting cascades.  Sources should be embedded in structured regions, filaments and galaxy clusters, whose scale is $l \sim$~Mpc that is comparable to the interaction length of UHE photons.  Observations suggest EGMFs with $\sim \mu \rm G$ in cluster centers and $\sim 0.1~\mu \rm G$ in cluster outskirts~\citep[e.g.,][]{CT02}.  Based on a physically motivated, turbulent dynamo model, sophisticated simulations suggested that filaments have $B_{\rm EG} \sim~{\rm a~few}\times 10$~nG~\cite{Ryu+08,CR09}.  In addition, magnetic fields with $\sim {10}^{1-2}$~nG are expected in haloes of old elliptical galaxies ($l \sim 1-3$~Mpc) or magnetized galactic winds from galaxies ($l \sim 0.5-1$~Mpc)~\citep[e.g.,][]{FL01,Ber+06}, and small scale fields could be even stronger.  In such magnetized regions, the synchrotron loss length, $\lambda_{\rm syn} \simeq 240~{\rm pc}~\gamma_{e,13.5}^{-1} B_{\rm EG,-7.5}^{-2}$, is much shorter than $l$ and the inverse-Compton (IC) loss length, $\lambda_{\rm IC}^{\rm KN} \sim 6.7~{\rm Mpc}~\gamma_{e,13.5} (10/[\ln (5.8 \times {10}^{4} \gamma_{e,13.5})-2])$ (in the Klein-Nishina regime), so UHE pairs are quickly depleted via synchrotron emission with typical energies,  
\begin{equation}
E_{\rm syn} \simeq 0.37~{\rm TeV}~\gamma_{e,13.5}^2 B_{\rm EG,-7.5}.
\end{equation} 
Here $\gamma_e = {10}^{13.5} \gamma_{e,13.5}$ and $B_{\rm EG} = {10}^{-7.5}~{\rm G}~B_{\rm EG,-7.5}$.  
Noticing that the energy fraction of UHE photons converted into UHE pairs is $\sim (1-e^{-l/\lambda_{\gamma \gamma}})$, the intrinsic synchrotron fluence (or flux for a steady source) at $E_{\rm syn} \approx 0.01 (\hbar e B_{\rm EG}/m_e^3 c^5) E_{p}^2$ is estimated to be 
\begin{equation}
E^2 \phi \sim \frac{1-e^{-l/\lambda{\gamma \gamma}}}{8 \pi d^2} 
\left( \frac{1}{2} f_{\rm esc} f_{p \gamma} \tilde{\mathcal E}_{\rm CR}^{\rm iso} \right), 
\end{equation}
where $\tilde{\mathcal E}_{\rm CR}^{\rm iso} \equiv E_p^2 \frac{d N_{\rm CR}^{\rm iso}}{d E_p}$ is the differential cosmic-ray energy input and $d$ is the distance.  The synchrotron spectrum of the structured region is roughly expressed as $E^2 \phi \propto E^{(\alpha-p)/2}$ (in the limit of $1-e^{-l/\lambda_{\gamma \gamma}} \sim l/\lambda_{\gamma \gamma}$).  
The deflection angle, $\theta_{\rm EG} \approx \sqrt{2} \lambda_{\rm syn}/ \sqrt{3} r_L \simeq 3.5 \times {10}^{-4}~\gamma_{e,13.5}^{-2} B_{\rm EG,-7.5}^{-1}$, is typically smaller than $\theta_j$, so high-energy emission is beamed.  The time spread is crucial for transient sources, and from Eq.~(2)  we get
\begin{equation}
{\Delta t}_{\rm EG} \approx \theta_{\rm EG}^2 (l/2c) \simeq 0.27~{\rm yr}~E_{\rm syn,11.5}^{-2} l_{\rm Mpc}
\end{equation}
for $l<\lambda_{\gamma \gamma}$, which suggests $\lesssim$~yr transients at $\gtrsim 0.1$~TeV.  

UHECRs themselves may leave the source, though details of the escape process are uncertain.  They can also make synchrotron $\gamma$ rays in the structured regions~\cite{GA05,Kot+11}.  As recently studied, structured EGMFs cause significant deflections and time delays~\citep[e.g.,][]{Tak+06,TM11}.  The deflection angle is $\theta_{\rm CR} \approx \sqrt{2 l \lambda_{\rm co}}/3 r_L \simeq 0.044 \bar{B}_{\rm EG,-8} {(\lambda_{\rm co}/l)}^{1/2} l_{\rm Mpc} E_{p,20}^{-1}$ for volume-averaged fields of $\bar{B}_{\rm EG} \sim 10$~nG in filaments~\cite{Ryu+08,CR09}, and the time spread due to the EGMF around the source,
\begin{equation}
{\Delta t}_{\rm CR} \approx \theta_{\rm CR}^2 (l/4c) \simeq 1.6 \times {10}^{3}~{\rm yr}~\bar{B}_{\rm EG,-8}^2 (\lambda_{\rm co}/l) l_{\rm Mpc}^3 E_{p,20}^{-2}, 
\end{equation}
is much longer than $\Delta t_{\rm EG}$, where $\lambda_{\rm co}$ is the coherent length of structured EGMFs.  Eq.~(5) also agrees with numerical calculations~\cite{TM11}.  Note that, though the volume filling factor is uncertain, such EGMFs imply effective fields of $B_{\rm eff} \lambda_{\rm eff}^{1/2} \sim 0.3~{\rm nG}~{\rm Mpc}^{1/2}$, consistent with upper limits from the Faraday rotation measure~\cite{KO11,TM11}.  
The total time spread ${\Delta T}_{\rm CR}$ can be longer due to the void EGMF and intervening structured EGMFs~\cite{TM11}.  

We are mainly interested in emissions from the structured regions, but emission may also come from the void region in which the EGMF is weaker and then the IC cascade will be developed~\citep[e.g.,][]{Mur+09,NV10,EAK11}.  Note that recent \textit{Fermi}-era limits on the void EGMF~\cite{EAK11} do not contradict much stronger EGMFs expected in clusters and filaments~\citep[e.g.,][]{KO11}.  For cascaded pairs in the Thomson regime, the IC loss length is $\lambda_{\rm IC} \simeq 72~{\rm kpc}~\gamma_{e,7}^{-1}$ and upscattered CMB photons have $E_{\rm IC} \approx (4/3) \gamma_e^2 \varepsilon_{\rm CMB} \simeq 88~{\rm GeV}~\gamma_{e,7}^2$.  The deflection angle is $\theta_{\rm IGV} \approx \sqrt{2} \lambda_{\rm IC}/ \sqrt{3} r_L \simeq 1.1 \gamma_{e,7}^{-2} B_{\rm IGV,-13}$, so the IC cascade emission in voids is diluted unless the EGMF is weak enough.  The EGMF dependence is more crucial for transients.  Noticing the effective loss length of UHE photons, $\lambda_{\gamma \gamma}^{\rm eff} \simeq 47~{\rm Mpc}~E_{\gamma,19}$~\cite{BS00}, one roughly obtains $\Delta t_{\rm IGV} \sim \theta_{\rm IGV}^2 (\lambda_{\gamma \gamma}^{\rm eff}/2 c) \simeq 7.1 \times {10}^{6}~{\rm yr}~E_{\rm IC,11.5}^{-2} B_{\rm IGV,-13}^2 \lambda_{\gamma \gamma,50~\rm Mpc}^{\rm eff}$, implying that such an IC pair echo is irrelevant if $B_{\rm IGV} \gtrsim {10}^{-16.5}$~G, similarly to cases for primary multi-TeV photons.

\begin{figure}[tb]
\includegraphics[width=\linewidth]{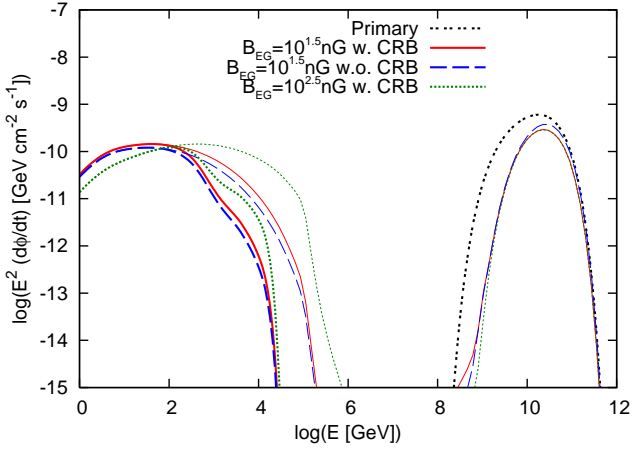}
\caption{
\small{$\gamma$-ray energy fluxes from structured regions embedding a steady AGN-like source at $z=0.15$, with an injection spectrum $\tilde{L}_{\gamma} = {10}^{44.2}~{\rm erg/s}~{(E/E_\gamma^{\rm max})}^{0.5} e^{-E_\gamma^{\rm min}/E} e^{-E/E_\gamma^{\rm max}}$, where $E_{\gamma}^{\rm max}={10}^{19.5}$~eV and $E_{\gamma}^{\rm min}={10}^{18.5}$~eV.  Here the jet opening angle $\theta_j=0.1$ is also used.  Thick/thin solid, dashed and dotted curves represent cases after/before $\gamma \gamma$ attenuation in the void region.
}}
\end{figure}  

\section{Results of cascade $\gamma$-ray spectra}
To calculate high-energy emission produced in intergalactic space, we solve Boltzmann equations in which $\gamma \gamma$ pair creation, synchrotron and IC emissions and adiabatic loss due to cosmic expansion are included~\cite{BS00,Mur09,Lee98}.  The energy and number conservation was monitored, and we focus on beamed emissions with $\theta_{\rm EG} (\gamma_e) < \theta_j$.  Based on observations and simulations~\cite{FL01,CT02,Ber+06,Ryu+08,CR09}, as in Takami \& Murase (2011), we consider structured regions of $l=2$~Mpc, with $B_{\rm EG}={10}^{1.5}$~nG (filament) or $B_{\rm EG}={10}^{-0.5}~\mu {\rm G}$ (galaxy cluster).  For the CIB, we use the low-IR model given by Kneiske et al. (2004).  For the uncertain CRB, we take the high CRB model in Protheroe \& Biermann (1996), but we also consider cases without the CRB. 

First, to see basic features of the emission from structured regions, we consider a steady source using an analytical injection spectrum (see Figure~2).  This case corresponds to an AGN with $\tilde{L}_{\rm CR}={10}^{46.5}$~erg/s, $f_{p \gamma} = 0.01 {(E_p/E_p^{\rm max})}^{0.5}$ and $E_p^{\rm max}={10}^{20.5}$~eV.  One sees that numerical spectra are consistent with analytical expectations, and the synchrotron emission from the structured regions is seen at $\sim (0.1-1)$~TeV.  
Also, our results are not very sensitive to the CRB unless $E_{p}^{\rm max} \gg {10}^{20.5}$~eV.  
Note that the TeV emission is observed as a point source since $l \theta_{\rm EG} \ll d$ (while the almost steady and unbeamed, synchrotron pair halo emission could be expected at $\lesssim$~GeV).

Next, we take GRBs as examples of transient sources.  For steady sources, the emission may be contaminated by other emissions from the source and runaway UHECRs, whereas it makes a unique signal for transients owing to $\delta T \ll \Delta t_{\rm EG} \ll \Delta t_{\rm CR}$, where $\delta T$ is the source duration.  To evaluate $f_{p \gamma}$ accurately and discuss detectability in more detail, we here calculate primary spectra of pionic $\gamma$ rays leaving the source using a detailed numerical code for the $p \gamma$ reaction~\cite{Mur07,Mur08}, where spectra of generated and runaway $\gamma$ rays are obtained, taking account of $\tau_{\gamma \gamma}$ in the emission region~\cite{Mur09}.  Calculations are first executed in the comoving frame, and then transformed to results in the observer frame.  
We assume the $E_p^{-2}$ spectrum of protons escaping from the acceleration zone, with given $E_p^{\rm max}$ (that is limited by $t_{\rm acc} < t_{\rm cool}$ and $t_{\rm dyn} < t_{\rm cool}$ for the parameter of the acceleration region, $\xi_{B,a} \equiv L_B/L_\gamma \sim 1$). 
    
In Figure~3, we show energy fluences of $\gamma$ rays for $\tilde{\mathcal E}_{\rm CR}^{\rm iso} = {10}^{53}$~erg (which is consistent with the UHECR energy budget $E_p^2 \frac{d \dot{N}_{\rm CR}}{d E_p} \sim {10}^{44}~{\rm erg}~{\rm Mpc}^{-3}~{\rm yr}^{-1}$ and the local HL GRB rate $\rho \sim 1~{\rm Gpc}^{-3}~{\rm yr}^{-1}$; Liang et al. 2007), with different strengths of the structured EGMF.  
Using $\Delta t_{\rm EG} (\gamma_e;E_\gamma) = (1+z) (1-\cos \theta_{\rm EG})~{\rm min}(l/c,\lambda_{\gamma \gamma}/c)$, we also calculate fluences of $\gamma$ rays received in $1$~yr, and we obtain $E^2 \phi \sim {10}^{-2.5}~{\rm GeV}~{\rm cm}^{-2}$ in the TeV range.  The fluxes averaged over $1$~yr are $\sim {10}^{-10}~{\rm GeV}~{\rm cm}^{-2}~{\rm s}^{-1}$ (and fluxes at earlier times are larger, $\sim {10}^{-9.5}~{\rm GeV}~{\rm cm}^{-2}~{\rm s}^{-1}$ at $0.1$~yr after the burst, and intrinsically harder), while the CTA sensitivity for integration time of 50~hr is $\sim {10}^{-10.5}~{\rm GeV}~{\rm cm}^{-2}~{\rm s}^{-1}$ at $\sim$~TeV~\cite{CTA10}, so synchrotron pair echoes from HL GRBs at $z \lesssim 0.3-0.4$ are detectable with follow-up observations.  The High Altitude Water Cherenkov experiment (HAWC) has lower sensitivity ($\sim 2 \times {10}^{-9}~{\rm GeV}~{\rm cm}^{-2}~{\rm s}^{-1}$ for one-year observation) but will also be helpful for closer bursts.  The results agree with analytical estimates, and higher-energy $\gamma$ rays have shorter time spreads while $\sim$~GeV $\gamma$ rays cannot be observed by \textit{Fermi} because they are suppressed by long time delays.  
Note that prompt emission from the source ceases after $\delta T \sim {10}^{1-2}$~s, whereas almost steady synchrotron emission induced by UHECRs escaping from the source is negligible since its flux ratio to the UHE-photon-induced synchrotron emission is $\lesssim f_{p \gamma}^{-1} (l/\lambda_{p \gamma}) ({\Delta t}_{\rm EG}/{\Delta t}_{\rm CR}) \sim 3 \times {10}^{-5} f_{p \gamma,-1}^{-1}  l_{\rm Mpc} ({\Delta t}_{\rm EG,0.3~\rm yr}/{\Delta t}_{\rm CR,\rm kyr}) $ (at $\sim 0.3$~TeV) for UHECR production with $\delta T<{\Delta t}_{\rm EG}$.  Here $\lambda_{p \gamma} \sim 100$~Mpc is the energy loss length of $\sim {10}^{20}$~eV protons.  

\begin{figure}[tb]
\includegraphics[width=\linewidth]{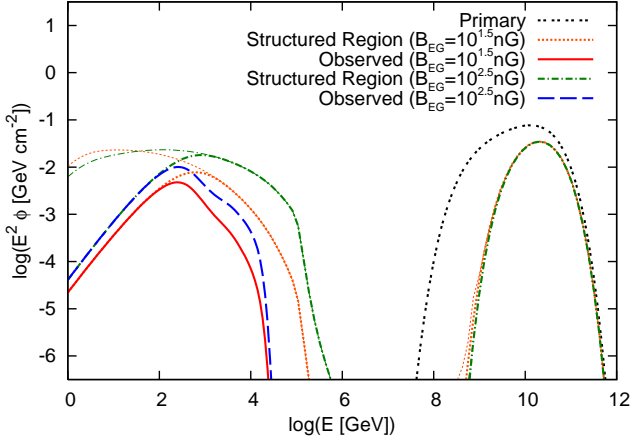}
\caption{
\small{
$\gamma$-ray energy fluences from structured regions embedding a HL GRB at $z=0.15$. The parameters of the emission region are: $L_\gamma^b={10}^{51.5}$~erg/s, $\Gamma={10}^{2.5}$, $r={10}^{14.5}$~cm, $\varepsilon_{\rm ob}^b=500$~keV, $\alpha=1$ and $2.2$, $\varepsilon_{\rm ob}^{\rm sa}=10$~eV, and $E_p^{\rm max}={10}^{20.5}$~eV are used.
Thick curves stand for contributions received in 1~yr, while thin curves for total energy fluences of beamed emissions.  The CRB is included.} 
 }
\end{figure}
\begin{figure}[tb]
\includegraphics[width=\linewidth]{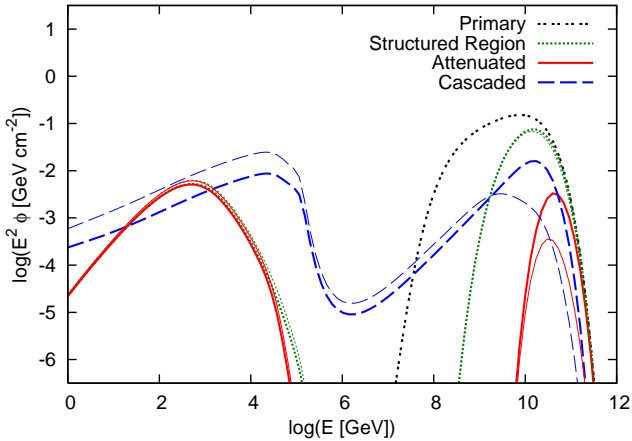}
\caption{
\small{$\gamma$-ray fluences from a structured region ($B_{\rm EG}={10}^{1.5}$~nG) embedding a LL GRB at $d=20$~Mpc.  The parameters of the emission region are: $L_\gamma^b={10}^{48}$~erg/s, $\Gamma={10}^{1.5}$, $r={10}^{15}$~cm, $\varepsilon_{\rm ob}^b=10$~keV, $\alpha=1$ and $2.2$, $\varepsilon_{\rm ob}^{\rm sa}={10}^{-1}$~eV, and $E_p^{\rm max}={10}^{20.2}$~eV are used~\citep[c.f.][]{Mur09}.  Thick/thin solid, dashed and dotted curves represent cases without/with the CRB.  Possible components of $\gamma$ rays cascaded in the void region are also shown. 
} 
}
\end{figure}

As suggested in Murase et al. (2006), low-luminosity (LL) GRBs such as GRB 060218 ($d \sim 140$~Mpc) and GRB 980425 ($d \sim 40$~Mpc) may be more important as local UHECR sources, since they are dimmer but more common than HL GRBs~\cite{Lia+07}.
In Figure~4, we show $\gamma$-ray fluences for $\tilde{\mathcal E}_{\rm CR}^{\rm iso}={10}^{50.5}$~erg (implied by their local rate, $\rho \sim {10}^{2.5}~{\rm Gpc}^{-3}~{\rm yr}^{-1}$; Liang et al. 2007).  
Expected fluences for a burst at 20~Mpc (e.g., in the Virgo cluster) are $E^2 \phi \sim {10}^{-2.5}~{\rm GeV}~{\rm cm}^{-2}$, so the synchrotron signal can be seen by CTA for nearby bursts as in the case shown in Figure~3.   

We have considered that $\gamma$ rays are attenuated in voids.  This is reasonable for GeV-TeV $\gamma$ rays from transients, since some lower limits ($B_{\rm IGV} \lesssim {10}^{-16}$~G) are strong enough~\citep[e.g.,][]{NV10} (and strong $B_{\rm eff}$ seems favored for HL GRBs to be UHECR sources; Murase \& Takami 2009), or the IC emission may be suppressed by isotropization via plasma instabilities.  But, we show the IC cascade for generality.  
Its flux ratio in the TeV range is $\sim {\Delta t}_{\rm EG}/{\Delta t}_{\rm IGV}$, so the IC pair echo from the void region can be important if $B_{\rm IGV} \lesssim {10}^{-16.5}$~G.  Also, the IC pair echo induced by runaway UHECRs is less significant, since its flux ratio to the UHE-photon-induced synchrotron pair echo is at most $\lesssim f_{p \gamma}^{-1} ({\Delta t}_{\rm EG}/{\Delta T}_{\rm CR}) \sim 3 \times {10}^{-3} f_{p \gamma,-1}^{-1}  ({\Delta t}_{\rm EG,0.3~\rm yr}/{\Delta T}_{\rm CR,\rm kyr})$ (depending on the uncertain escaping fraction and beaming of UHECRs), which will even be smaller when $B_{\rm IGV} \gtrsim {10}^{-15}$~G.

Note that time spread of cascade UHE photons is much shorter, $\Delta t_{\rm IGV}^{\rm KN} \approx ({\mathcal N}/3) (\lambda_{\rm IC}^{\rm KN} \lambda_{\rm co} d/18 r_L^2 c) \simeq 11~{\rm s}~{\mathcal N}_1 \gamma_{e,13.5}^{-1} (B_{\rm IGV,-15}^2 \lambda_{\rm co,Mpc}) d_{20~\rm Mpc}$, where ${\mathcal N} \sim d/(\lambda_{\gamma \gamma}+\lambda_{\rm IC})$, so they may be detected for nearby transient sources~\cite{Mur09}.  (The diffuse UHE photon background is dominated by cosmogenic UHE photons produced by runaway UHECRs, due to $f_{p \gamma} < 1$, and consistent with the PAO limits.)  Results are also shown in Figure~4, taking account of escape from the structured region (while Murase 2009 simply assumed that $\sim 1/2$ of $\gamma$ rays are lost).   
UHE photons and accompanied UHE pairs may pass another structured region, including the local supercluster, and/or our Galaxy, where we expect further synchrotron pair echo emission, which enhances detectability.

\section{Implications and discussions}
We studied effects of the magnetized environment embedding UHECR accelerators, and then showed that UHE photon beams inevitably lead to synchrotron emission.  In particular, any UHE $\gamma$-ray burst or flare should produce a characteristic echo with $\Delta t_{\rm EG} \sim 0.1-1~{\rm yr} \times {\rm min}[l_{\rm Mpc},\lambda_{\gamma \gamma, \rm Mpc}] \ll \Delta t_{\rm CR}$ in the TeV range, from the immediate environment, galaxy groups and clusters, filaments, and all corresponding sites surrounding the source.  Note that our results are conservative in that the fluxes can be higher if $\gamma$ rays are attenuated by possible additional photon fields in the immediate environment.  The signal is detectable by future HAWC and CTA.  For the echo, observations with other wavelengths and messengers are relevant to trigger coordinated search (especially for CTA) and estimate source parameters.  Then detecting a $0.1-1$~yr transient in the TeV range may suggest the accelerator in the magnetized regions with $l \gtrsim$~Mpc.  Observing synchrotron spectra and light curves can also provide invaluable information on the EGMFs as well as properties of the sources. 
Importantly, the signal is useful to identify \textit{distant extragalactic accelerators} beyond $\sim 100$~Mpc, and it will favor powerful $E_p \gtrsim {10}^{20}$~eV proton accelerators over heavy ion sources~\citep[e.g.,][]{Gor+08,Mur+08,Wan+08}.  It is possible to consider the Galactic transient origin for heavy nuclei around $10$~EeV, but the emissions from their extragalactic counterparts are weak~\cite{CKN10}.  

Synchrotron pair echo emission induced by runaway UHE photons from transients becomes more prominent than other emissions, after $\delta T$ (for source emission including hadronic and leptonic $\gamma$ rays), since $\Delta t_{\rm EG}$ is shorter than $\Delta t_{\rm CR}$ (for synchrotron emission induced by runaway UHECRs).  
It is applicable to UHECR production associated with AGN flares ($\delta T \sim {10}^{4-5}$~s), GRB prompt emission and afterglows.  Though GRB afterglow emission may last longer, its flux declines with time as $\propto t^{-1.2}$ after $\sim {10}^{3}$~s and $\propto t^{-2.2}$ after the jet break time of $\sim$~day~\cite{Zha07}, so the pair echo can be dominant and the temporal behaviors are distinguishable.  For steady sources, discrimination would be harder, but the source emission will be more variable and the almost steady UHECR-induced emission may be less relevant if most UHECRs are isotropized or cooled before they escape.  

Discriminating between the synchrotron and IC pair echoes would also be possible.  The synchrotron spectrum typically has $E^2 \phi \propto E^{(\alpha-2)/2}$ as distinct from the hard IC cascade spectrum (see Figure~4), $E^2 \phi \propto E^{1/2}$ for $E \lesssim 10~{\rm TeV}~E_{\rm cut,14}^2$ (where $E_{\rm cut}$ is the cutoff by the CMB or CIB).  Also, $\Delta t_{\rm EG}$ is generally different from $\Delta t_{\rm IGV}$. 

 
\begin{acknowledgements}
K.M. thanks the referee, J. Beacom, S. Inoue, K. Ioka, K. Kotera, H. Takami and B. Zhang for discussion, and support by JSPS and CCAPP.
\end{acknowledgements}


\begin{thebibliography}{}
\bibitem[Abbasi et al. 2010a]{HiR10a}
Abbasi, R.U., et al. 2010a, ApJ, 713, L64
\bibitem[Abbasi et al. 2010b]{HiR10b}
---. 2010b, Phys. Rev. Lett., 104, 161101
\bibitem[Abraham et al. 2007]{PAO07}
Abraham, J., et al. 2007, Science, 318, 938
\bibitem[Abraham et al. 2010a]{PAO10a}
---. 2010a, Astropart. Phys., 34, 314 
\bibitem[Abraham et al. 2010b]{PAO10b}
---. 2010b, Phys. Rev. Lett., 104, 091101
\bibitem[Abreu et al. 2011]{PAO11}
Abreu, P., et al. 2011, arXiv:1107.4805
\bibitem[Atoyan \& Dermer 2001]{AD01}
Atoyan, A., \& Dermer, C.D. 2001, Phys. Rev. Lett., 87, 221102
\bibitem[Atoyan \& Dermer 2003]{AD03}
---. 2003, ApJ, 586, 79
 
\bibitem[Bertone et al. 2006]{Ber+06}
Bertone, S., Vogt, C., \& En{\ss}lin, T. 2006, MNRAS, 370, 319 
\bibitem[Bhattacharjee \& Sigl 2000]{BS00}
Bhattacharjee, P., \& Sigl, G. 2000, Phys. Rep., 327, 109 
\bibitem[Biermann \& de Souza 2011]{BS11}
Biermann, P.L., \& de Souza, V. 2011, arXiv:1106.0625
\bibitem[Blandford 2000]{Bla00}
Blandford, R.D. 2000, Phys. Scr. T85, 191

\bibitem[Calvez et al. 2010]{CKN10}
Calvez, A., Kusenko, A., \& Nagataki, S. 2010, Phys. Rev. Lett., 105, 091101
\bibitem[Carilli \& Taylor 2002]{CT02}
Carilli, C.L., \& Taylor, G.B. 2002, ARA\&A, 40, 319
\bibitem[Cho \& Ryu 2009]{CR09}
Cho, J., \& Ryu, D. 2009, ApJ, 705, L90 
\bibitem[CTA Consortium 2010]{CTA10}
CTA Consortium 2010, arXiv:1008.3703 

\bibitem[Dermer 2007]{Der07}
Dermer, C.D. 2007, arXiv:0711.2804 
\bibitem[Dermer et al. 2009]{Der+09}
Dermer, C.D., et al. 2009, New J. Phys., 11, 065016 

\bibitem[Essey et al. 2011]{EAK11}
Essey, W., Ando, S., \& Kusenko, A. 2011, Astropart. Phys., 35, 135

\bibitem[Farrar \& Gruzinov 2009]{FG09}
Farrar, G.R., \& Gruzinov, A. 2009, ApJ, 693, 329 
\bibitem[Furlanetto \& Loeb 2001]{FL01}
Furlanetto, S.R., \& Loeb, A. 2001, ApJ, 556, 619
 
\bibitem[Gabici \& Aharonian 2005]{GA05}
Gabici, S., \& Aharonian, F.A. 2005, Phys. Rev. Lett. 95, 251102
\bibitem[Gorbunov et al. 2008]{Gor+08}
Gorbunov, D., et al. 2008, JETP Lett., 87, 461
\bibitem[Granot et al. 2008]{Gra+08}
Granot, J., Cohen-Tanugi, J., \& Do Couto E Silva, E. 2008, ApJ, 677, 92

\bibitem[Kneiske et al. 2004]{Kne+04}
Kneiske, T.M., et al. 2004, A\&A, 413, 807
\bibitem[Kotera \& Lemoine 2008]{KL08}
Kotera, K., \& Lemoine, M. 2008, Phys. Rev. D, 77, 123003  
\bibitem[Kotera \& Olinto 2011]{KO11}
Kotera, K., \& Olinto, A.V. 2011, ARA\&A, 49, 119
\bibitem[Kotera et al. 2011]{Kot+11}
Kotera, K., Allard, D., \& Lemoine, M. 2011, A\&A, 527, A54 

\bibitem[Lee 1998]{Lee98}
Lee, S. 1998, Phys. Rev. D, 58, 043004
\bibitem[Lemoine \& Waxman 2009]{LW09}
Lemoine, M., \& Waxman, E. 2009, J. Cosmol. Astropart. Phys., 11, 009
\bibitem[Li \& Waxman 2007]{LW07}
Li, Z., \& Waxman, E. 2007, arXiv:0711.4969
\bibitem[Liang et al. 2007]{Lia+07}
Liang, E., et al. 2007, ApJ, 662, 1111 

\bibitem[Murase 2007]{Mur07}
Murase, K. 2007, Phys. Rev. D, 76, 123001
\bibitem[Murase 2008]{Mur08}
---. 2008, Phys. Rev. D, 78, 101302(R)
\bibitem[Murase 2009]{Mur09}
---. 2009, Phys. Rev. Lett., 103, 081102
\bibitem[Murase \& Takami 2009]{MT09}
Murase, K., \& Takami, H. 2009, ApJ, 690, L14
\bibitem[Murase et al. 2006]{Mur+06}
Murase, K., Ioka, K., Nagataki, S., \& Nakamura, T. 2006, ApJ, 651, L5
\bibitem[Murase et al. 2008]{Mur+08}
---. 2008, Phys. Rev. D, 78, 023005
\bibitem[Murase et al. 2009]{Mur+09}
Murase, K., Zhang, B., Takahashi, K., \& Nagataki, S. 2009, MNRAS, 396, 1825

\bibitem[Neronov \& Vovk 2010]{NV10}
Neronov, A., \& Vovk, I. 2010, Science, 328, 73
 
\bibitem[Protheroe \& Biermann 1996]{PB96}
Protheroe, R.J., \& Biermann, P.L. 1996, Astropart. Phys., 6, 45
\bibitem[Rachen \& M\'esz\'aros 1998]{RM98}
Rachen, J.P., \& M\'esz\'aros, P. 1998, Phys. Rev. D, 58, 123005
\bibitem[Razzaque et al. 2004]{Raz+04}
Razzaque, S., M\'esz\'aros, P., \& Zhang, B. 2004, ApJ, 613, 1072
\bibitem[Ryu et al. 2008]{Ryu+08}
Ryu, D., et al. 2008, Science, 320, 909

\bibitem[Takami \& Murase 2011]{TM11}
Takami, H., \& Murase, K. 2011, arXiv:1110.3245
\bibitem[Takami et al. 2006]{Tak+06}
Takami, H., Yoshiguchi, H., \& Sato, K. 2006, ApJ, 639, 803

\bibitem[Wang et al. 2008]{Wan+08}
Wang, X.Y., Razzaque, S., \& M\'esz\'aros, P. 2008, ApJ, 677, 432
\bibitem[Waxman \& Bahcall 1997]{WB97}
Waxman, E., \& Bahcall, J. 1997, Phys. Rev. Lett., 78, 2292 

\bibitem[Zaw et al. 2009]{Zaw+09}
Zaw, I., Farrar, G.R., \& Greene, J. 2009, ApJ, 696, 1218
\bibitem[Zhang 2007]{Zha07}
Zhang, B. 2007, Chin. J. Astron. Astrophys., 7, 1
\end{thebibliography}
\end{document}